\documentclass{kluwer}    
\def\degr{^{\circ}}
\newdisplay{guess}{Conjecture}
\begin{document}
\begin{article}
\begin{opening}
\title{Magnetic fields in our Galaxy: How much do we know?}
\subtitle{I. Disk fields within a few kpc of the Sun}
\author{J.L. \surname{Han}}
\runningauthor{J.L. Han}
\runningtitle{Magnetic fields in our Galaxy}
\institute{Beijing Astronomical Observatory,
             CAS, Jia-20 DaTun Rd., ChaoYang District,
             Beijing 100012, China (hjl@bao.ac.cn)
              }
\date{July 5, 2000. Astrophysics \& Space Science, in press (Proceedings
of IAUC 182)}

\begin{abstract}
The large scale magnetic fields of our Galaxy have been mostly revealed
by rotation measures ($RM$s) of pulsars and extragalactic radio sources.
In the disk of our Galaxy, the average field strength over a few
kpc scale is about 1.8 $\mu$G, while the total field, including the
random fields on smaller scales, has a strength of about 5 $\mu$G.
The local regular field, if it is part of the large scale field of a
bisymmetric form, has a pitch angle of about $-8\degr$. There are at least
three, and perhaps five, field reversals from the Norma arm to the outer skirt
of our Galaxy.  
\end{abstract}
\end{opening}           

\section{Introduction}

The large scale structure of the magnetic fields in the Milky Way Galaxy
is difficult to figure out as it can never be measured completely.
Many efforts in the last decade have resulted in a good deal of knowledge,
although far from enough. A closely-related topic is the measurement
of the magnetic fields in nearby spiral galaxies, mainly by multi-wavelength
radio polarization observations (see review by Beck et al. 1996). The regular
magnetic field generally has similar orientations parallel to the spiral
arms in galactic disks. The magnetic structure could be a superposition
of the large scale structure with some local features. The fields
were probably generated by dynamos.

I will summarize results on the local field in the vicinity of the Sun,
mainly the regular field within a few kpc. I have ignored
the measurements of fields in individual clouds which are connected to,
but are not part of, the large scale fields. Limited by space here, the
magnetic field in the Galactic halo (Han et al. 1997; 1999:
HMQ99) and the Galactic center will be discussed in a companion paper.

There are three key parameters to describe the local field: 1) the field
directions, or practically, the pitch
angle, $p$, of the field, which is the deviation of the field direction from
$l=90\degr$; 2) the field strength of
the regular and irregular components, as well as the total field strength,
and 3) the locations of field reversals. Although we concentrate on
the regular field, the irregular random field is the overwhelmingly
dominant component in the interstellar medium.

\section{Pitch angle}

Using the largest dataset of 7500 stars, Andreasyan \& Makarov (1989)
concluded that the local field is concentrated in the spiral arms.
Re-analyzing the dataset of 7000
stars of Mathewson \& Ford (1970), Heiles (1996) obtained for the pitch
angle $p=-7.2\degr \pm 4.1\degr$ (see sign definition of the pitch
angle in HMQ99) and concluded that the field lines almost follow the
spiral pattern. These optical data, all very local and mostly within 1 kpc,
have been ultimately used already.

Pulsars probably are the only probes of the Galactic magnetic fields
on larger scales. Extensive studies became possible only after the $RM$s
of a large sample of pulsars had been measured. Hamilton \& Lyne (1987)
obtained the $RM$s of 185 pulsars. Rand \& Kulkarni (1989: RK89)
obtained a field direction of $p=6\degr\pm4\degr$ from $RM$s of 116
pulsars within 3 kpc, a result probably affected by selection effects.
Han \& Qiao (1994: HQ94) found
$p=-8.2\degr\pm0.5\degr$ from model-fitting to the
data of carefully selected pulsars within 3 kpc. This result was confirmed by
Indrani \& Deshpande (1998: ID98) and is consistent with the value from
optical data (Heiles 1996). The projected $RM$ distribution of disk pulsars
within $-15\degr < l < 15\degr$ shows the transition from positive to
negative values almost exactly at the galactic longitude expected from
this pitch angle (HMQ99). So, we conclude that the pitch angle of the
local regular field is $p = -8\degr$, with a maximum uncertainty of
$2\degr$.

\vspace{-2mm}
\section{Field strength}

E.M. Berkhuijsen (Fig.~5 in Beck et al.\ 1996), using the deconvolved surface
brightness of synchrotron emission from our Galaxy, estimated the total
field strength to range from about 10 $\mu$G at a Galactic radius of $R_0=4$ 
kpc to about 5 $\mu$G in the vicinity of the Sun. 

In fact, the average field strength in our Galaxy can be {\it directly}
determined from pulsar $DM$ and $RM$, $\langle B_{||} \rangle =1.232 \; 
RM/DM$, the best approach to get observational values of the large scale
magnetic fields. Manchester (1974) was the first to get the strength of the
local regular field, $B = 2.2\pm0.4 \mu$G.  Thomson \& Nelson (1980: TN80)
found an average field of $\sim 1\mu$G and a maximum ordered field of 3.5
$\mu$G. Lyne \& Smith (1989) used a large dataset of pulsar $RM$s and $DM$s
and found the local field strength of about $2\sim3 \mu$G. RK89
found that the average strength of the local field is about
 $1.6\pm 0.2 \mu$G, and the strength of the random field is about $5 \mu$G.
Ohno \& Shibata (1993) confirmed this random field strength and
they emphasized that the value is independent of the cell size.
Rand \& Lyne (1994) suggested that the total field strength
perhaps reaches $6\sim7 \mu$G at about $R_0 = 6$ kpc
and probably increases smoothly towards the Galactic center.
HQ94 obtained from pulsar data for the field in the
vicinity of the Sun, a regular field strength of about 1.4$\mu$G,
with a maximum regular field strength (in the field reversal model)
of about $1.8\pm0.3 \mu$G. They also concluded from the residual after
model-fitting
that the magnetic field energy stored in the random component is 3.7 times
that in the regular field, which is consistent with the estimate later
made by Zweibel \& Heiles (1997) and implies that the total field is
about $5\mu$G.
ID98 confirmed all these results. Both HQ94 and ID98 showed that the
regular field is stronger in the interarm region.

All the above results reach a consensus: the strength of the
regular field is $1.8\pm0.3\mu$G and total field is about 5$\mu$G
locally, but probably becomes stronger towards the Galactic
center.

\vspace{-2mm}
\section{Field reversals}

There is no doubt that field reversals exist in our Galaxy. The key
points are, 1) how many reversals are there, and 2) where do they occur?
Evidence for the nearest field reversal, about 0.2 kpc towards the Galactic
center and near the Carina-Sagittarius arm, was first found by TN80
from model-fitting to pulsar $RM$ data and by Simard-Normandin
\& Kronberg (1980: SK80) from model-fitting to the $RM$s of extragalactic
radio sources (EGRS). All research thereafter (RK89;
Clegg et al. 1992; HQ94; ID98) confirmed this field reversal by
similar or different use of more pulsar $RM$ data.

Field reversals, beyond and near the Perseus arm, were first revealed
by Lyne \& Smith (1989) by comparing the $RM$ values of distant pulsars
with those of extragalactic radio sources. The field reversal
near the Perseus arm was suggested by the bi-symmetric spiral (BSS) model
(SK80; Sofue \& Fujimoto 1983: SF83; HQ94) and was
confirmed by Clegg et al. (1992) using newly determined $RM$s of EGRS and
by HMQ99 using the $RM$s of pulsars and EGRS. There are some
indications (HMQ99) of a further field reversal beyond
the Perseus arm at about 15 kpc from the Sun.

The second field reversal in the inner Galaxy, located near the
Crux-Scutum arm, was first shown by $RM$ data in Rand \& Lyne (1994).
Such a field was also predicted by the BSS model (SF83;
HQ94) and later confirmed by more pulsar
$RM$ data near $l\sim 327\degr$ (HMQ99).
Marginal evidence was found for a field reversal near the Norma arm
(HMQ99). More $RM$ data from pulsars, newly discovered in the Parkes
multibeam survey will help to figure out the field there.

In summary, there are at least three, perhaps five, field reversals known
in our Galaxy. The reversed fields are separated by spiral arms.

\section{Concluding remarks}

There are three models for the global structure of magnetic fields
in the disc of our Galaxy. RK89 and Rand \& Lyne (1994) argued that
pulsar $RM$s are consistent with a concentric-ring model for the field.
This can work for the zero-order approximation, but the pitch angle
for the local field of $\sim -8\degr$ does not support this model.
Vall\'ee (1996) argued that the field has an axi-symmetric spiral form
and no field
reversal was allowed beyond $R_0 = 8$ kpc. The fact of at least one
field reversal near or beyond the Perseus arm is not consistent with
this model. Early analyses of the $RM$ distribution of extragalactic
radio sources (SK80; SF83) suggested that the Galactic magnetic field
has a BSS form, in which the field direction reverses from arm to arm.
This model seems to be supported by the statistical study of pulsar $RM$s
(HQ94; ID98; HMQ99) and is consistent with the pitch angle and the number of
field reversals located over a large range of galactic radii.

\smallskip
{\small I am very grateful to many colleagues, especially, Prof. R.N.
Manchester and Prof. G.J. Qiao, for working together with me for a long
time to improve the knowlege on the magnetic fields of our Galaxy which
I presented here. The research of the author is supported by the National
Natural Science Foundation of China, the National Major Project of Fundamental
Research, and the Su-Shu Huang Research Foundation of CAS.}

\end{article}
\end{document}